\documentclass{ws-procs975x65}          
\usepackage{amssymb,amsmath}
\usepackage{bm}

\def\ANT{\nearrow\hspace{-11pt}A}

\def\a{\alpha}
\def\b{\beta}
\def\g{\gamma}
\def\vt{\vartheta}
\def\d{\delta}
\def\vta{\vartheta}
\def\ve{\varepsilon}
\def\sq2{\sqrt{\frac{\varepsilon_0}{\mu_0}} }  
\def\c{\gamma}

\def\d{\delta}
 \def\m{\mu}
\def\n{\nu}

\begin{document}

\title{Premetric approach in gravity and electrodynamics}

\author{Yuri N. Obukhov}

\address{Russian Academy of Sciences, Nuclear Safety Institute (IBRAE),\\
B. Tulskaya 52, 115191 Moscow, Russia\\
E-mail: obukhov@ibrae.ac.ru}

\begin{abstract}
The basics of the premetric approach are discussed, including the essential details of the formalism and some of its beautiful consequences. We demonstrate how the classical electrodynamics can be developed without a metric in a quite straightforward way: Maxwell's equations, together with the general response law for material media, admit a consistent premetric formulation. Furthermore, we show that in relativistic theories of gravity, the premetric program leads to a better understanding of the interdependence between topological, affine, and metric concepts.
\end{abstract}


\bodymatter

\section{Development of premetric ideas}

The study of the metric-free, or premetric, models has a rich and long history. The corresponding timeline of works which developed the premetric approach in physics is as follows. As a starting point, we mention H. Minkowski (1908) who established the special-relativistic (Poincar\'e-covariant) formalism for Maxwell's theory. F. Kottler (1912) subsequently provided the generally covariant formulation of electromagnetism which was eventually taken by A. Einstein (1916) as a basis for a quasi-premetric formulation of Maxwell's theory. F. Kottler (1922) then actually pioneered the premetric approach by constructing, based on integral conservation laws, metric-free theory for Newton's gravity and Maxwell's electrodynamics. Similar line was followed by \'E. Cartan (1923) in electrodynamics and by D. van Dantzig (1934) who proposed a general premetric program in physics. Important technical contributions came from I. Tamm (1925) who studied a general linear constitutive law and from A. Sommerfeld (1948) who coined, following Mie, the notions of extensive and intensive quantities. The formal structure of electrodynamics was thoroughly investigated by E.J. Post (1962) and by C. Truesdell and R. Toupin (1960) in the framework of the formal field theory. An important issue of recovering the metric (deriving light-cone) was then clarified in the works of R. Toupin (1965), M. Sch\"onberg (1971), and A. Jadzcyk (1979). W.-T. Ni (1973) was the first to propose an axion-dilaton extension of Maxwell's theory, and subsequently F. Wilczek (1987) looked into experimental issues of axion electrodynamics. See\cite{birk,PT1,PT2,PT3} for exact references. 

Why going metric-free? Quoting Edmund Whittaker (1953): {\it Since the notion of metric is a complicated one, which requires measurements with clocks and scales, generally with rigid bodies, which themselves are systems of great complexity, it seems undesirable to take metric as fundamental, particularly for phenomena which are simpler and actually independent of it}.
Furthermore, as we know (in Einstein's approach), metric is identified with the gravitational field. Thereby, one can say that a metric formulation of a physical theory is ``contaminated'' by gravity. Hence, by revealing the metric-free relations one actually finds the most fundamental physical structures. In this sense, premetric approach helps to understand the essential interdependence between topological, affine, and metric concepts. Moreover, from the experimental standpoint, counting is obviously the simplest measurement!

\subsection{Principles of premetric approach}

Premetric axiomatics is straightforward. Construction of a metric-free physical model is inherently based on conservation laws. The latter should be phenomenologically verified by means of a certain counting process without use of the metric. Next, all physical objects are divided into two sets: extensive variables ({\it how much?}) are distinguished from intensive variables ({\it how strong?}). Based on the conservation laws, the two sets of {\it fundamental equations} then naturally arise for excitations (extensities) and for strengths (intensities). Final step: to convert the model into a predictive theory, one adds the {\it linking equations}, or constitutive relations, between excitations and field strengths.

The essence of the premetric art is formulated as follows: Fundamental equations are metric-free; the metric only enters via linking equations (constitutive relations).

\section{Premetric electrodynamics}

Classical electrodynamics admits a consistent premetric formulation\cite{birk}. Based on the conservation of electric charge $dJ = 0$, one arrives at $dH = J$, and the magnetic flux conservation results in $dF = 0$. Conservation laws require just {\it counting procedures} -- of electric charges and magnetic flux lines. No distance concept is needed, therefore the premetric framework arises naturally. 

Excitations $H = ({\mathcal D}, {\mathcal H})$ are measurable via the {\it charge}. These are extensive variables (how much?). Field strengths $F = (E, B)$ are measurable via the {\it force}. They are intensive variables (how strong?). 

{\it Fundamental equations} of electrodynamics are metric-free:
\begin{equation}
dH = J,\qquad dF = 0.\label{max}
\end{equation}

 {\it Linking equations} (constitutive relations) yield a predictive theory
\begin{equation}\label{lin}
H = \kappa\left[F\right],\qquad H_{\alpha\beta} = \kappa_{\alpha\beta}{}^{\mu\nu}F_{\mu\nu}.  
\end{equation}
The metric is hidden/encoded in the constitutive tensor $\kappa_{\alpha\beta}{}^{\mu\nu}$. Constitutive relation can be more general -- nonlinear and even nonlocal. 

Using the local coordinates $x^i = (t, x^a)$, we have the $(1+3)$-decompositions
\begin{equation}
H = −\,{\mathcal H}_adx^a\wedge dt + {\mathcal D}^a\epsilon_a,\qquad
F = E_adx^a\wedge dt + B^a\epsilon_a,\label{HF} 
\end{equation}
(with $dx^a$ and $\epsilon_a$ bases of spatial 1- and 2-forms) and the local and linear constitutive law (\ref{lin}) is recast into
\begin{eqnarray}\label{lin13}
{\mathcal H}_a = -\,{\mathfrak C}^{b}{}_a\,E_b + {\mathfrak B}_{ba}\,B^b\,,\qquad
{\mathcal D}^a = -\,{\mathfrak A}^{ba}\,E_b + {\mathfrak D}_{b}{}^a\,B^b\,.
\end{eqnarray}
With $\varepsilon^{ab}=\varepsilon^{ba}$, $\mu^{-1}_{ab}=\mu^{-1}_{ba}$, $\gamma^c{}_c=0$, a convenient parametrization reads as\cite{birk}:
\begin{align}
{\mathfrak A}^{ba} &= -\,\varepsilon^{ab} +\epsilon^{abc}n_c, &
{\mathfrak B}_{ba} &= \mu^{-1}_{ab} - \epsilon_{abc}m^c,\\ 
{\mathfrak C}^{b}{}_a &= \gamma^{b}{}_a - s_a{}^b +
\delta_a^b\,s_c{}^c + \alpha\,\delta_a^b, & 
{\mathfrak D}_{b}{}^a &= \gamma^{a}{}_b + s_b{}^a -\delta_b^a\,s_c{}^c 
+ \alpha\,\delta_b^a.
\end{align} 
Thereby the fine structure of the constitutive tensor is eventually revealed:
\begin{eqnarray}\label{fine}
 \kappa = \underbrace{\left(\begin{array}{cc}
        \gamma^b{}_a & \mu_{ab}^{-1} \\ -\varepsilon^{ab} &
        \gamma^a{}_b \end{array}\right)}_{ principal\>{\rm part\> 20\>
      comp.}} + \underbrace{ \left(\begin{array}{cc} - s_a{}^b
        +\delta_a^b s_c{}^c &\>\; - \epsilon_{abc}m^c \\ 
        \epsilon^{abc}n_c & s_b{}^a - \delta_b^a s_c{}^c
  \end{array}\right)}_{ skewon\> {\rm part\> 15\> comp.}}
+ \underbrace{\alpha\left(\begin{array}{cc}\delta_a^b&0\\
 0&\delta_b^a\end{array}\right)}_{ axion\>{\rm part\> 1 \>comp.}}
\end{eqnarray}

\section{Premetric formulation of gravity}

Kottler in 1922 formulated a premetric nonrelativistic (Newton's) gravity. However, the spacetime metric$\,=\,$gravity in general relativity (GR) theory, and one may ask whether a premetric {\it relativistic} gravity makes any sense? In GR, the answer is negative. However, a {\it teleparallel} framework offers a viable opportunity\cite{PT1,PT2,PT3}. Qualitatively, one proceeds by replacing the electric charge with the ``gravitational charge'' $=$ mass (energy-momentum). The conservation of the gravitational charge $d\Sigma_\alpha = 0$ yields $dH_\alpha = \Sigma_\alpha$, thereby introducing the gravitational excitation 2-form $H_\alpha = {\frac 12}H_{ij\alpha} dx^i\wedge dx^j = {\frac 12}\check{H}^{\rho\sigma}{}_\alpha\epsilon_{\rho\sigma}$. Furthermore, the gravitational flux conservation results in $dF^\alpha = 0$, introducing the gravitational field strength 2-form $F^\alpha = {\frac 12}F_{ij}{}^\alpha dx^i\wedge dx^j =  d\vartheta^\alpha$, so that the coframe $\vartheta^\alpha$ plays a role of the gravitational potential. The premetric formulation of gravity can be then consistently constructed along the lines of premetric electromagnetism. The corresponding gravitational-electromagnetic analogy is summarized in Table \ref{table}.

\begin{table}
\tbl{Premetric electromagnetism-gravity analogy}
{\begin{tabular}{@{}ccc@{}}
  \toprule
  Objects and Laws & Electromagnetism & Gravity  \\
 \colrule
  Source current 3-form & $J$& $\Sigma_\a$\\
  Conservation law  &$dJ=0$&$d\Sigma_\a=0$\\
  Excitation 2-form &$H$&$H_\a$\\
  Inhomogeneous field equation &$dH=J$&$dH_\a = \Sigma_\a
  ={}^{(\vartheta)}\Sigma_\a+{}^{\text{(m)}}\Sigma_\a$\\
  Field strength 2-form &$F$&$F^\a$\\
  Homogeneous field equation &$dF=0$&$dF^\a=0$\\
  Potential 1-form &$A$&$\vt^\a$\\
  Potential equation &$dA=F$&$d\vt^\a=F^\a$\\
  Lorentz force &$f_\a=\left(e_\a\rfloor F\right) \wedge J$&$f_\a
=\left(e_\a\rfloor F^\b\right) \wedge {}^{\text{(m)}}\Sigma_\b$\\
  Energy-momentum 3-form & 
  ${\frac 12}(F\wedge e_\a\rfloor H - H\wedge e_\a\rfloor F)$
  & ${\frac 12}(F^\beta\wedge e_\a\rfloor H_\beta - H_\beta\wedge e_\a\rfloor F^\beta)$ \\
  Lagrangian 4-form &$\Lambda=- {\frac 12} F\wedge H$&$\Lambda=- {\frac 12} F^\a\wedge H_\a$\\
  Constitutive tensor &$\chi^{\a\b\g\d}$&$\chi^{\b\g}{}_{\a}{}^{\nu\rho}{}_\mu$ \\
  \botrule
\end{tabular}}\label{table}
\end{table}

\subsection{Gravitational constitutive tensor}

The local and linear constitutive law
\begin{equation}\label{linG}
\check{H}^{\a\b}{}_\mu = {\frac 12}\chi^{\a\b}{}_{\mu}{}^{\rho\sigma}{}_\nu \,F_{\rho\sigma}{}^\nu
\end{equation}
describes a rich class of gravity models. The symmetry properties $\chi^{\b\g}{}_{\a}{}^{\nu\rho}{}_\mu  = -\,\chi^{\g\b}{}_{\a}{}^{\nu\rho}{}_\mu  = -\,\chi^{\b\g}{}_{\a}{}^{\rho\nu}{}_\mu$ yield a simple number count $(6\times 4)\times(6\times 4) = 576$ of components of the constitutive tensor. The constitutive relation is {\it reversible} when $\chi^{\b\g}{}_{\alpha}{}^{\nu\rho}{}_{\mu} = \chi^{\nu\rho}{}_{\mu}{}^{\b\g}{}_{\alpha}$, with a reduced number of components: $24(24+1)/2 = 300$. 

Splitting $\chi{}^{\alpha\beta}{}_\mu{}^{\gamma\delta}{}_\nu = \overset{+}{\chi}{}^{\alpha\beta}{}_\mu{}^{\gamma\delta}{}_\nu  + \overset{-}{\chi}{}^{\alpha\beta}{}_\mu{}^{\gamma\delta}{}_\nu$ into reversible and irreversible parts,
\begin{equation}
  \overset{+}{\chi}{}^{\alpha\beta}{}_\mu{}^{\gamma\delta}{}_\nu 
  := \frac12( \chi{}^{\alpha\beta}{}_\mu{}^{\gamma\delta}{}_\nu 
  + \chi{}^{\gamma\delta}{}_\nu{}^{\alpha\beta}{}_\mu ),\quad
  \overset{-}{\chi}{}^{\alpha\beta}{}_\mu{}^{\gamma\delta}{}_\nu 
  := \frac12( \chi{}^{\alpha\beta}{}_\mu{}^{\gamma\delta}{}_\nu 
    - \chi{}^{\gamma\delta}{}_\nu{}^{\alpha\beta}{}_\mu),
\end{equation}
one derives a decomposition $\chi{}^{\alpha\beta}{}_\mu{}^{\gamma\delta}{}_\nu = \sum\limits_{I=1}^8 {}^{[I]}\chi{}^{\alpha\beta}{}_\mu{}^{\gamma\delta}{}_\nu$ of the constitutive tensor into six irreducible parts: 
\begin{align}
  {}^{[1]}\chi{}^{\alpha\beta}{}_\mu{}^{\gamma\delta}{}_\nu & 
= \overset{+}{\chi}{}^{\alpha\beta}{}_{(\mu}{}^{\gamma\delta}{}_{\nu)} 
- \overset{+}{\chi}{}^{[\alpha\beta}{}_{(\mu}{}^{\gamma\delta]}{}_{\nu)} , 
 & {}^{[2]}\chi{}^{\alpha\beta}{}_\mu{}^{\gamma\delta}{}_\nu & 
=\overset{-}{\chi}{}^{\alpha\beta}{}_{[\mu}{}^{\gamma\delta}{}_{\nu]} 
- \overset{-}{\chi}{}^{[\alpha\beta}{}_{[\mu}{}^{\gamma\delta]}{}_{\nu]} , \\
  {}^{[3]}\chi{}^{\alpha\beta}{}_\mu{}^{\gamma\delta}{}_\nu & 
= \overset{-}{\chi}{}^{\alpha\beta}{}_{(\mu}{}^{\gamma\delta}{}_{\nu)} , 
 & {}^{[4]}\chi{}^{\alpha\beta}{}_\mu{}^{\gamma\delta}{}_\nu & 
= \overset{+}{\chi}{}^{\alpha\beta}{}_{[\mu}{}^{\gamma\delta}{}_{\nu]} , \\
  {}^{[5]}\chi{}^{\alpha\beta}{}_\mu{}^{\gamma\delta}{}_\nu & 
=  \overset{+}{\chi}{}^{[\alpha\beta}{}_{(\mu}{}^{\gamma\delta]}{}_{\nu)}, 
 & {}^{[6]}\chi{}^{\alpha\beta}{}_\mu{}^{\gamma\delta}{}_\nu & 
=  \overset{-}{\chi}{}^{[\alpha\beta}{}_{[\mu}{}^{\gamma\delta]}{}_{\nu]}.
\end{align}
We call ${}^{[1]}\chi$ a reversible symmetric principal part (principal-1), ${}^{[4]}\chi$ a reversible antisymmetric principal part (principal-2), ${}^{[5]}\chi$ a reversible axion (axion-1), ${}^{[2]}\chi$ a skewon antisymmetric principal part (skewon-1), ${}^{[3]}\chi$ a skewon symmetric principal part (skewon-2), and ${}^{[6]}\chi$ a skewon axion (axion-2). 

Comparing to electromagnetism, we have three parts in Maxwell's theory (\ref{fine}): $\chi^{\mu\nu\a\b} = {\frac 12}\epsilon^{\mu\nu\rho\sigma}\kappa_{\rho\sigma}{}^{\a\b} = {}^{(1)}\chi^{\mu\nu\a\b} + {}^{(2)}\chi^{\mu\nu\a\b} + {}^{(3)}\chi^{\mu\nu\a\b}$: the principal ${}^{(1)}\chi$ piece and the axion ${}^{(3)}\chi$ (both reversible); and the skewon ${}^{(2)}\chi$ (irreversible).

When the metric exists on spacetime, the most general linear constitutive tensor
\begin{eqnarray}
\chi^{\a\b}{}_{\m}{}^{\g\d}{}_{\n}(g) = {\frac{1}{\varkappa}}\left[\b_1
\,g^{\g[\a}g^{\b]\d}g_{\m\n} +\,\b_2\,\d^{[\a}_{\,\,\m}g^{\b][\g}\d^{\d]}_{\n} 
+ \b_3\,\d^{[\a}_{\,\,\n}  g^{\b][\g}\d_{\m}^{\d]}\right.\\
\left. + \,{\beta}_4\,\ve^{\a\b\g\d}\,g_{\m\n} + {\beta}_5\,
  \ve^{\a\b[\g}{}_{[\m}\,\delta^{\d]}_{\n]} 
 + \b_6\left(\delta_{(\m}^{[\a}\ve^{\b]\g\d}{}_{\n)} 
  - \delta_{(\m}^{[\g}\ve^{\d]\a\b}{}_{\n)}\right)\right]\label{linM}
\end{eqnarray}
encompasses four irreducible parts: principal-1 proportional to $\b_1$ and $(\b_2 + \b_3)$, principal-2 proportional to $\b_5$ and $(\b_2 - \b_3)$, skewon-2 $\sim\b_6$, and axion-1 $\sim\b_4$.

\subsection{Propagation of gravitational waves}

In a geometric optics approximation, $F^\alpha = {\cal F}^\alpha e^{i\Phi}$ with rapidly varying phase $\Phi$ and slowly changing amplitude ${\cal F}^\a$ (eikonal ansatz). Alternatively, in Hadamard's theory, wave is described as a discontinuity across characteristic hypersurface (wave front). In any case, vacuum field equations reduce to algebraic system for amplitudes
\begin{equation}\label{geo}
dH_\alpha = 0,\ dF^\alpha = 0\qquad\Longrightarrow\qquad\chi^{\m\n}{}_{\a}{}^{\rho\sigma}{}_\b\,
{\cal F}_{\rho\sigma}{}^\b\,q_\n = 0,\ \epsilon^{\mu\nu\rho\sigma}\,{\cal F}_{\rho\sigma}{}^\a\,q_\nu = 0,
\end{equation}
with the wave covector $q = d\Phi = q_\mu\vta^\mu$. As a result, ${\cal F}_{\rho\sigma}{}^\a = A_\rho{}^\a q_\sigma - A_\sigma{}^\a q_\rho$, and from (\ref{geo}) for the amplitude $A_\m{}^\nu$ we find a {\it characteristic equation}
\begin{equation}\label{chA}
M^\mu{}_\alpha{}^\nu{}_\beta\,A_\nu{}^\b = 0\,,\qquad
M^\mu{}_{\alpha}{}^\nu{}_\beta := \chi^{\mu\rho}{}\!_{\a}{}^{\nu\sigma}{}\!_\b\,q_\rho\,q_\sigma .
\end{equation}
The dispersion relation (Fresnel equation) arises as a solvability condition of (\ref{chA}).
Decomposing $A_\a{}^\b = \ANT_\a{}^\b + A\d^\b_\a$ into traceless part ($\ANT_\a{}^\a = 0$) and trace $A = {\frac 14}A_\c{}^\c$:
\begin{equation}
N^\mu{}_\alpha{}^\nu{}_\beta\!\ANT_\nu{}^\b = 0,\qquad N^\mu{}_\alpha{}^\nu{}_\beta = M^\mu{}_\alpha{}^\nu{}_\beta M^\rho{}_\rho{}^\sigma{}_\sigma - M^\mu{}_\alpha{}^\rho{}_\rho M^\sigma{}_\sigma{}^\nu{}_\beta .
\end{equation}
Notice the gauge freedom: $A_\nu{}^\b\rightarrow A_\nu{}^\b + q_\nu C^\b$ leaves (\ref{chA}) invariant for any $C^\b$.

In the metric-dependent case (\ref{linM}), for generic class of models with $2\beta_1 - \beta_2 - \beta_3 \neq 0$ and $2\beta_1 + \beta_3 \neq 0$ the characteristic equation splits into decoupled equations
\begin{align}
(2\beta_1 - \beta_3)\left\{3\left(q^2\delta^\mu_\a 
- q^\mu q_\a\right)(q^2\delta^\nu_\b - q^\nu q_\b) 
    - (q^2g_{\a\b} - q_\a q_\b) q^\mu q^\nu\right\}\ANT_{(\mu\nu)} = 0,&\\
   (2\beta_1 + \beta_3)\left(q^2\delta^\mu_\a - q^\mu q_\a\right)
    (q^2\delta^\nu_\b - q^\nu q_\b)\!\!\ANT_{[\mu\nu]} = 0,&
\end{align}
and the scalar mode is recovered from $3q^2A = \ANT_\nu{}^\mu q_\mu q^\nu$. Therefore, only symmetric $\ANT_{(\mu\nu)}$ mode is dynamical if $2\beta_1 = -\,\beta_3$, whereas only antisymmetric $\ANT_{[\mu\nu]}$ mode remains dynamical when $2\beta_1 = \beta_3$.

The teleparallel equivalent general relativity model GR$_{||}$ is a special case when both $2\beta_1 - \beta_2 - \beta_3 = 0$ and $2\beta_1 + \beta_3 = 0$. Explicitly: $\b_4 = \b_5 = \b_6 = 0$ and
\begin{equation}
\b_1 = -\,1\,,\qquad \b_2 = -\,4\,,\qquad \b_3 = 2.\label{GRE} 
\end{equation}
For GR$_{||}$ coupling constants (\ref{GRE}), the characteristic equation (\ref{chA}) reduces to
\begin{eqnarray}
q^2h_{\a\b} - q_\a q^\gamma h_{\gamma\b} - q_\b q^\gamma h_{\a\gamma} = 0,
\end{eqnarray}
which describes the usual spin 2 graviton mode $h_{\a\b} 
= A_{(\a\b)} - {\frac 12}g_{\a\b}A_\gamma{}^\gamma$.

\subsection{Recovering General Relativity}

The teleparallel equivalent GR$_{||}$ theory (\ref{GRE}) is distinguished among other coframe models by the following remarkable properties:
\begin{itemize}
\item  Field equations are invariant under local Lorentz group\cite{PP,Cho}\\
$\vta^\alpha\longrightarrow L^\alpha{}_\beta(x)\vta^\beta,\quad L^\alpha{}_\beta \in SO(1,3)$.
\item Black hole solutions exist\cite{OP}.
\end{itemize}
The gravitational Lagrangian of GR$_{||}$ actually reduces to Hilbert's Lagrangian $\Lambda = - {\frac 18}\chi^{\beta\gamma}{}_\alpha{}^{\rho\sigma}{}_\mu F_{\beta\gamma}{}^\alpha F_{\rho\sigma}{}^\m = - {\frac 1{2\varkappa}}\,R\,+\,$total derivative. The constitutive tensor has only two nontrivial irreducible parts -- principal-1 and principal-2.

{\it Premetric gravity summarized.} Fundamental equations are metric-free:
\begin{equation}
dH_\alpha = \Sigma_\alpha,\qquad dF^\alpha = 0.
\end{equation}
Metric enters only in linking (constitutive law) equations (\ref{linG}), where\cite{Guz} for ${\rm GR}_{||}$:
\begin{equation}
\chi^{\a\b}{}_{\mu}{}^{\g\d}{}_{\nu}= \frac{1}{\varkappa}\left(-g^{\g[\a}g^{\b]\d}g_{\mu\nu}
- 4\d^{[\a}_{\,\,\mu}g^{\b][\g}\d^{\d]}_{\nu} + 2\d^{[\a}_{\,\,\nu} g^{\b][\g}\d_{\mu}^{\d]}\right).
\end{equation}

\section{Conclusions and outlook: On Kottler's path}

Following the pioneering contributions by Kottler, Cartan and van Dantzig, the premetric program works universally in physics, embracing classical particle mechanics, kinetic theory, and (most importantly) electromagnetism and gravity. A feasible premetric gravity theory can be constructed on the basis of the energy-momentum conservation law as a teleparallel coframe model. The corresponding general local and linear constitutive relation encompasses {\sl six} irreducible parts (2 principal, 2 skewon, and 2 axion).

In premetric approach, one can view constitutive tensor $\chi^{\a\b}{}_{\mu}{}^{\g\d}{}_{\nu}$ as an independent variable. This opens new perspectives in electromagnetism and gravity theory such as:
\begin{itemize}
\item Natural extensions with {\it axion, skewon, and dilaton} fields\cite{Ni}
\item Natural extension to {\it parity odd} contributions\cite{PP,MH}
\item Convenient framework to discuss {\it Lorentz violating} models\cite{K}
\item Nonlinear gravity models, $f(R)$ and $f(T)$, in particular\cite{Cap,Hoh}
\item Nonlocal constitutive laws\cite{Mash,non}
\end{itemize}

\section*{Acknowledgments}

This talk is essentially based on the papers\cite{PT1,PT2,PT3} written in collaboration with Friedrich Hehl, Yakov Itin and Jens Boos. 
I am grateful to Marcus Werner for the kind invitation and support which made it possible for me to attend MG15.

\end{document}